\documentclass{LT23auth}

 \usepackage{graphicx}

\begin{document}


\begin{frontmatter}



\title{Nuclear Magnetic Relaxation Rate in the Vortex State of a Chiral
$p$-Wave Superconductor}


\author[okayama]{Nobuhiko Hayashi\corauthref{cor1}}
\ead{hayashi@mp.okayama-u.ac.jp}
and
\author[tokyo]{Yusuke Kato}

\address[okayama]{Computer Center, Okayama University,
Okayama 700-8530, Japan}
\address[tokyo]{Department of Basic Science, University of Tokyo,
Tokyo 153-8902, Japan}

\corauth[cor1]{Corresponding author. Fax: +81-86-251-7244}

\begin{abstract}
   The site-selective
nuclear spin-lattice relaxation rate $T_1^{-1}$
is theoretically studied
inside a vortex core in a chiral $p$-wave superconductor
within the framework of the quasiclassical theory of superconductivity.
   It is found that $T_1^{-1}$ at the vortex center
depends on the sense of the chirality
relative to the sense of the magnetic field.
   Our numerical result
shows a characteristic difference in $T_1^{-1}$
between the two chiral states,
$\bar{k}_x + i \bar{k}_y$ and $\bar{k}_x - i \bar{k}_y$
under the magnetic field.
\end{abstract}

\end{frontmatter}

   Site-selective nuclear magnetic resonance (NMR) method
was recently revealed to be
a powerful tool for experimentally investigating
the electronic structure inside vortex cores
in the mixed state of type-II superconductors
(see Refs.\ \cite{Takigawa99,Matsuda02,Mitrovic01}
and references cited therein).
   The site-selective NMR at the impurity site
was also proposed theoretically \cite{Matsumoto01}.
   The NMR technique as a probe of the electronic structure
with spatial resolution
is expected to reveal pairing symmetry
of unconventional superconductors,
because in spatially inhomogeneous systems there appear
properties specific to the unconventional superconductivity.

   The vortex core is one of such inhomogeneous
superconducting systems.
   In this paper, we study
the site-selective
nuclear spin-lattice relaxation rate $T_1^{-1}$
inside the vortex core in a chiral $p$-wave superconductor
with an unconventional pairing \cite{Sigrist99}
${\bf d}=\bar{\bf z}(\bar{k}_{x} \pm i \bar{k}_{y})$.
   We find that $T_1^{-1}(T)$ exhibits a characteristic
chirality dependence
as a function of the temperature $T$,
which might be experimentally observed as a sign of
the chiral pairing state.

   The system is assumed to be a two dimensional conduction layer
perpendicular to the magnetic field applied along
the $z$ axis.
   From now on, the notations are the same as those in Ref.\ \cite{Hayashi02}.
   We consider
the quasiclassical Green function
%
\begin{equation}
{\hat g}(i\omega_n,{\bf r},{\bar{\bf k}})=
-i\pi
\pmatrix{
g &
if \cr
-if^{\dagger} &
-g \cr
},
\label{eq:qcg}
\end{equation}
which is the solution of the Eilenberger equation \cite{serene},
\begin{eqnarray}
i v_{\rm F} {\bar{\bf k}} \cdot
{\bf \nabla}{\hat g}
+ \bigl[ i\omega_n {\hat \tau}_{3}-{\hat \Delta},
{\hat g} \bigr]
=0.
\label{eq:eilen}
\end{eqnarray}

   From the spin-spin correlation function \cite{Takigawa99},
we obtain the expression for $T_1^{-1}$
in terms of ${\hat g}$,
\begin{eqnarray}
\frac{T_1^{-1}(T)}{T_1^{-1}(T_\mathrm{c})}
& = &
\frac{1}{4T_\mathrm{c}}
\int^{\infty}_{-\infty}
\frac{\mathrm{d} \omega}{\cosh^{2}(\omega / 2T)}
W(\omega, -\omega),
%
%
\label{eq:T1}
\end{eqnarray}
\begin{eqnarray}
W(\omega, \omega')
=
\langle a_{11}(\omega) \rangle
\langle a_{22}(\omega') \rangle
-
\langle a_{12}(\omega) \rangle
\langle a_{21}(\omega') \rangle,
\label{eq:T1-w}
\end{eqnarray}
where the spectral function $\hat{a}(\omega,{\bf r},{\bar{\bf k}})
=\bigl(a_{ij}\bigr)$ is given as
\begin{eqnarray}
\hat{a}(\omega,{\bf r},{\bar{\bf k}})
 = \frac{- i}{2 \pi} \hat{\tau}_3 \bigl[
&  &
{\hat g}(i\omega_n \rightarrow \omega -i\eta ,
{\bf r},{\bar{\bf k}})                             \nonumber \\
&  & \mbox{} -
{\hat g}(i\omega_n \rightarrow \omega +i\eta ,{\bf r},{\bar{\bf k}})
\bigr],
\label{eq:spectral}
\end{eqnarray}
the symbol $\langle \cdots \rangle$
represents
the average over the Fermi surface,
and
$\eta$ is a small positive constant roughly representing
the impurity effect.

   Substituting vortex-center solution ${\hat g}$ \cite{Hayashi02}
based on the so-called zero-core vortex model \cite{Thuneberg84}
into Eq.\ (\ref{eq:T1-w}),
we obtain analytical expressions for $T_1^{-1}$ at the vortex center
in the case of each chiral state, $\bar{k}_{x} \pm i \bar{k}_{y}$.

\begin{figure}
\begin{center}
\small{(a)}
\includegraphics[width=5.54cm]{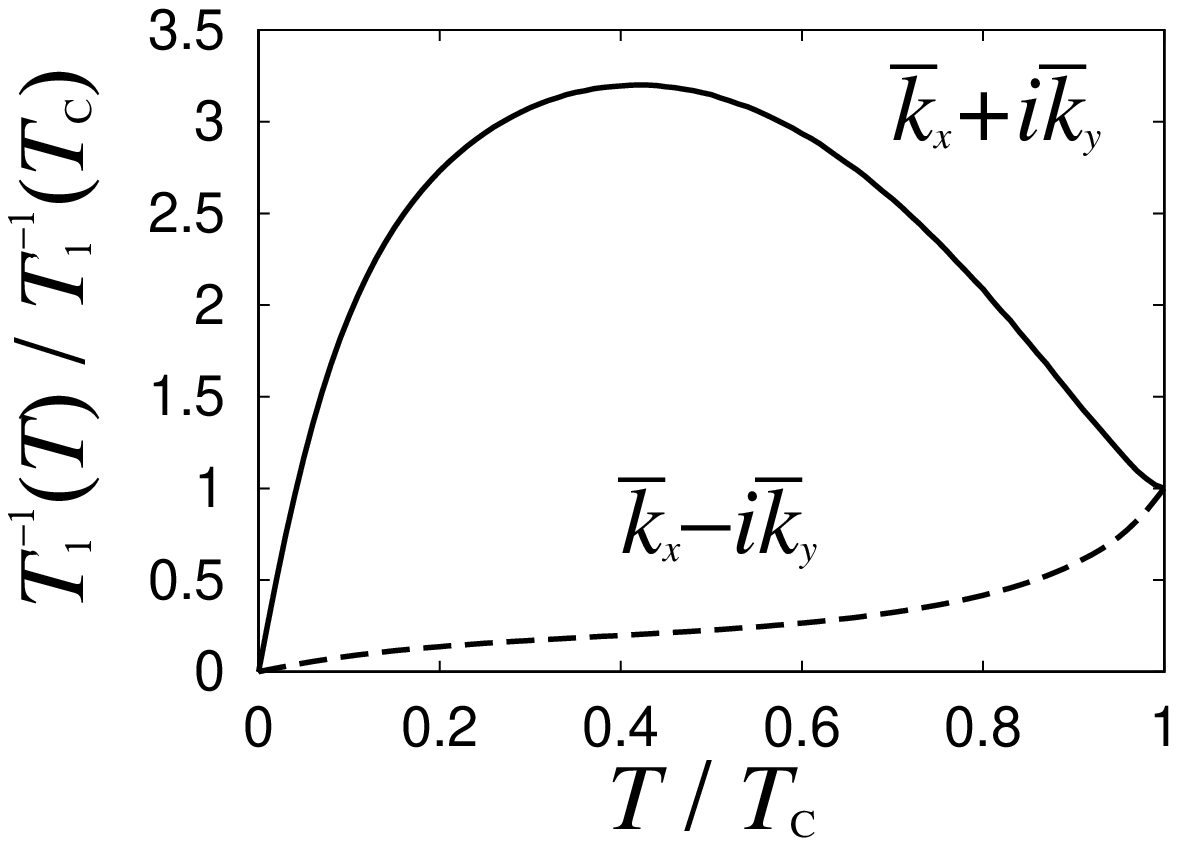}
\end{center}
\begin{center}
\small{(b)}
\includegraphics[width=5.54cm]{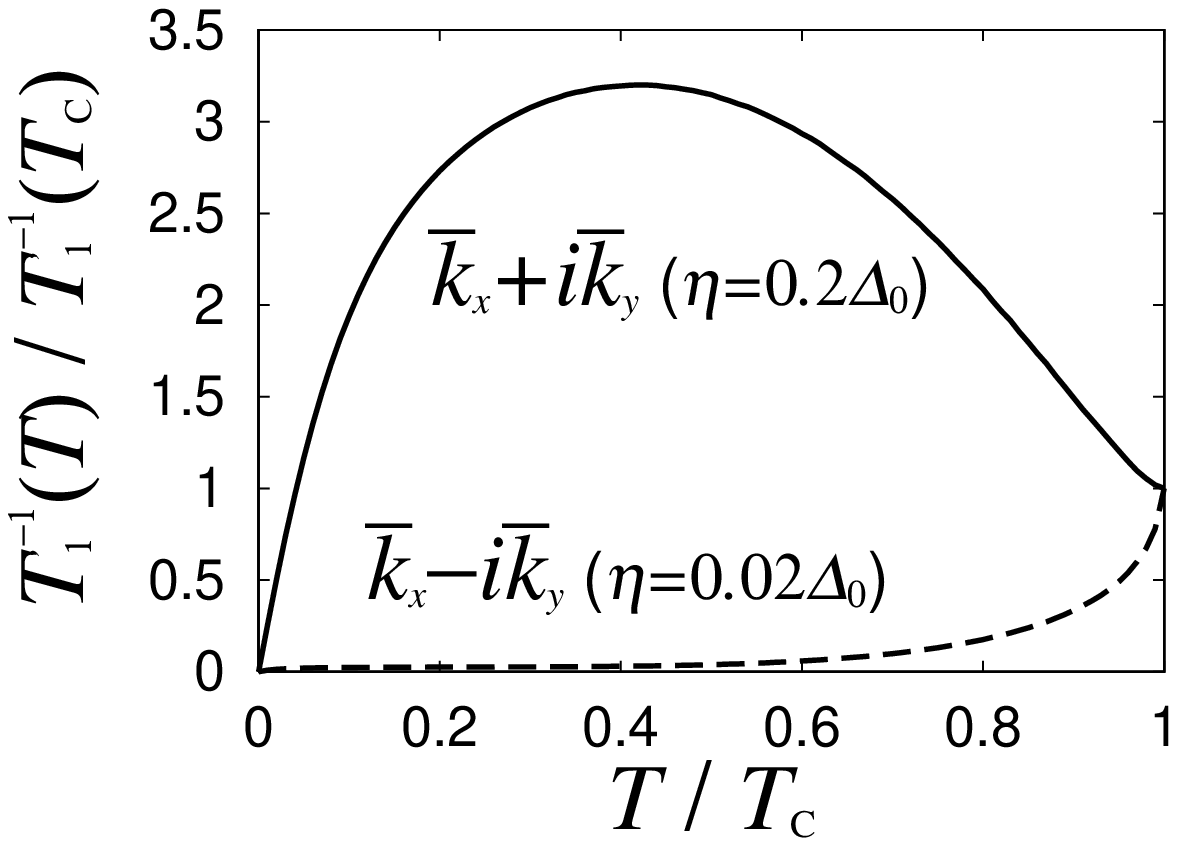}
\end{center}
\caption{
   $T_1^{-1}$ vs $T$ at the vortex center.
   The magnetic field is applied
in positive direction of the $z$ axis.
   The result for the $\bar{k}_{x} + i \bar{k}_{y}$ state
is identical to that for the $s$-wave pairing state.
   The parameter $\eta$
represents the smearing effect of the impurities.
   (a) $\eta = 0.2 \Delta_0$
($\Delta_0$ is the gap amplitude at $T=0$).
   (b) $\eta$ is set as $\eta=0.02 \Delta_0$
only for the $\bar{k}_{x} - i \bar{k}_{y}$ state (see text).
}
\label{fig:1}
\end{figure}

   In Fig.\ \ref{fig:1}(a), we show the result obtained by
numerically integrating
those expressions
with isotropic Fermi surface.
   It is noticeable that
$T_1^{-1}$ of the $\bar{k}_{x} - i \bar{k}_{y}$ state
is quite different from that of the $\bar{k}_{x} + i \bar{k}_{y}$ state.
   Note that the magnetic field is applied
in positive direction of the $z$ axis.
   In the $\bar{k}_{x} - i \bar{k}_{y}$ state,
$T_1^{-1}(T)$ is suppressed in wide $T$ region.
   This is because the second term in Eq.\ (\ref{eq:T1-w})
composed of the anomalous Green functions $f$ and $f^{\dagger}$
(or $a_{12}$ and $a_{21}$)
is nonzero at the vortex center \cite{Hayashi02}
and has minus contribution to $T_1^{-1}$
in the $\bar{k}_{x} - i \bar{k}_{y}$ state,
while it is zero in the other state.

    It was pointed out in Ref.\ \cite{Kato00} that
the impurity scattering rate inside the vortex core
of the $\bar{k}_{x} - i \bar{k}_{y}$ state
was one order smaller
than that of the $\bar{k}_{x} + i \bar{k}_{y}$ state.
   In an actual situation,
therefore,
such $T_1^{-1}$ as plotted in Fig.\ \ref{fig:1}(b) is anticipated.
   Here, $T_1^{-1}$ is calculated with smaller $\eta$
for the $\bar{k}_{x} - i \bar{k}_{y}$ state.
   It is noted that
$T_1^{-1}$ almost vanishes
in the $\bar{k}_{x} - i \bar{k}_{y}$ state.


   We note that our present result is
in contrast to a corresponding theoretical result for $T_1^{-1}(T)$
of Ref.\ \cite{Takigawa02-2}
obtained in the quantum limit
($k_{\rm F}\xi \sim 1$).
   In the result of Ref.\ \cite{Takigawa02-2},
there is not
such suppression of $T_1^{-1}(T)$ as seen in our Fig.\ \ref{fig:1}.
   A reasonable origin of this difference is as follows.
The calculations of $T_1^{-1}$ in
Ref.\ \cite{Takigawa02-2} are in the quantum limit
($k_{\rm F}\xi \sim 1$)
where the energy spectrum (the diagonal part of $\hat{a}$)
inside the vortex core is
quantized and
it dominantly determines $T_1^{-1}(T)$
as pointed out in Ref.\ \cite{Takigawa02-2},
while we base our calculations on
the quasiclassical theory relevant in
the opposite limit $k_{\rm F} \xi \gg 1$
where the vortex core spectrum is continuous
and
the coherence factor in Eq.\ (\ref{eq:T1-w})
(especially for the off-diagonal part of $\hat{a}$)
determines $T_1^{-1}(T)$.

   The chiral superconductivity is expected to be realized in
a material Sr$_2$RuO$_4$ \cite{Sigrist99,Maeno}.
   The result of this paper (Fig.\ \ref{fig:1}),
i.e., the chirality dependence of $T_1^{-1}$,
might be observed experimentally
by the site-selective NMR method \cite{Matsuda02,Mitrovic01}
in Sr$_2$RuO$_4$.
and such observations are expected to
be helpful to identify the pairing symmetry
in this material.
   Unfortunately,
it is certainly difficult to perform such a NMR experiment
because
the upper critical field is too small to attain
the relevant resonance frequency of NMR
in the case of applying the magnetic field
along the $c$($z$) axis perpendicular to the conduction layers
of Sr$_2$RuO$_4$.
   However, the upper critical field is larger
in the case of applying the magnetic field along the conduction layers
of this material, and therefore
the NMR experiments might be possible by
applying the magnetic field inclined to the $c$ axis in Sr$_2$RuO$_4$.
   To theoretically predict detailed behavior of $T_1^{-1}$
for the inclined magnetic fields,
further analysis will be needed.
   Such an analysis is left for future studies.



We thank M.\ Ichioka, M.\ Takigawa, M.\ Matsumoto, K.\ Machida,
N.\ Schopohl, and M.\ Sigrist
for helpful discussions.

\vspace{-0.5cm}



\begin{thebibliography}{00}




\bibitem{Takigawa99}
M. Takigawa {\it et al.},
Phys. Rev. Lett. {\bf 83} (1999) 3057; \\
J. Phys. Soc. Jpn. {\bf 69} (2000) 3943.

\bibitem{Matsuda02}
K. Kakuyanagi {\it et al.},
Phys. Rev. B {\bf 65} (2002) 060503(R).

\bibitem{Mitrovic01}
V. F. Mitrovi\'c {\it et al.},
Nature {\bf 413} (2001) 501.

\bibitem{Matsumoto01}
M. Matsumoto,
J. Phys. Soc. Jpn. {\bf 70} (2001) 2505.

\bibitem{Sigrist99}
M. Sigrist {\it et al.},
Physica C {\bf 317-318} (1999) 134; \\
A. G. Lebed, N. Hayashi,
Physica C {\bf 341-348} (2000) 1677.


\bibitem{Hayashi02}
N. Hayashi, Y. Kato,
Physica C {\bf 367} (2002) 41; \\
Phys. Rev. B {\bf 66} (2002) 132511.

\bibitem{serene}
J. W. Serene, D. Rainer,
Phys. Rep. {\bf 101} (1983) 221.

\bibitem{Thuneberg84}
E. V. Thuneberg {\it et al.},
Phys. Rev. B {\bf 29} (1984) 3913.

\bibitem{Kato00}
Y. Kato,
J. Phys. Soc. Jpn. {\bf 69} (2000) 3378; \\
Y. Kato, N. Hayashi,
J. Phys. Soc. Jpn. {\bf 70} (2001) 3368; \\
J. Phys. Soc. Jpn. {\bf 71} (2002) 1721.


\bibitem{Takigawa02-2}
M. Takigawa {\it et al.},
J. Phys. Chem. Solids {\bf 63} (2002) 1333.

\bibitem{Maeno}
Y. Maeno {\it et al.},
Nature {\bf 372} (1994) 532.


\end{thebibliography}
\end{document}